\documentclass[twocolumn,showpacs,preprintnumbers,amsmath,amssymb,showkeys]{revtex4}

\usepackage{graphicx}% Include figure files
\usepackage{dcolumn}% Align table columns on decimal point
\usepackage{bm}% bold math
\usepackage{bezier}

\begin{document}

\title{Photo-assisted heat current and Peltier coefficient in a metal/dot/metal junction}
\author{A. Cr\'epieux}
\affiliation{Centre de Physique Th\'eorique, CNRS,  Aix-Marseille Universit\'e, 163 avenue de Luminy, 13288 Marseille, France}

\date{\today}

\begin{abstract}

The photo-assisted heat current through a metal/dot/metal junction and its associated Peltier coefficient are computed in the framework of the time-dependent out-of-equilibrium Keldysh formalism in the presence of a dot energy modulation. When the frequency of the modulation is much larger than the amplitude of the modulation, the heat current follows the sinusoidal time evolution of the dot energy. This is no longer the case when the modulation frequency becomes of the order or smaller than the amplitude of the modulation. To characterize this non sinusoidal behavior, we have calculated the harmonics of the photo-assisted heat current. The zero-order harmonic can be expressed as an infinite sum of dc heat currents associated to a dot with shifted energies. It exhibits a devil staircase profile with non horizontal steps whereas it is established that the steps are horizontal for the zero-order harmonic of the photo-assisted electric current. This particularity is related to the fact that the dot heat is not a conserved quantity due to energy dissipation within the tunnel barriers.
 
\end{abstract}
\maketitle

\section{Introduction}

The photo-assisted electric current through an insulating barrier is the topic of a large number of studies both theoretically\cite{tien63,tucker85,lesovik94,crepieux03} and experimentally\cite{dayem62,cook67,kouwenhoven94,kozhevnikov00} in a wide range of systems, such as for example: quantum dots, normal or superconducting tunnel junctions, and Luttinger liquids. In the presence of a voltage modulation, the electrons can emit or absorb photons when they travel across the barrier and the resulting current is thus a superposition of dc currents. When the current-voltage characteristic is non-linear, it leads to a specific behavior for the current.\cite{cook67}

This is only recently that the study of photo-assisted heat current has emerged. On the theoretical side, the heat flow generated by an adiabatic quantum pump through a mesoscopic sample has been studied\cite{moskalets02,moskalets04}, as well as the photo-assisted heat flow in a normal metal/superconductor junction\cite{kopnin08}. Non-adiabatic pumping heat in an asymmetric double quantum dot has also been considered\cite{rey07}. Even more recently, the chiral heat transport in driven quantum Hall edges states\cite{arrachea11} and the microwave-mediated heat transport and thermoelectric effect in a quantum dot in the presence of Coulomb interaction\cite{chi12} have been calculated. On the experimental side, measurements of Seebeck voltage has been performed in magnetic tunnel junctions in the presence of a frequency modulated laser used to heat up the device\cite{walter11}.

The formalism that has been developed\cite{crepieux11} to calculate the time-dependent heat current with the help of out-of-equilibrium Green's functions allows to treat any type of time-dependent voltages, in particular a modulated one. It is therefore the appropriate formalism that has to be used when one considers time-dependent excitation in the quantum regime. It enables not only to calculate the photo-assisted electric and heat currents but also, as done here for the first time, the photo-assisted Peltier coefficient since this last quantity is defined as the ratio between the heat current and the electric current in the absence of temperature gradient. However, this formalism does not make possible the calculation of the photo-assisted Seebeck coefficient. Indeed, whereas this quantity is generally measured in an open circuit, it is not straightforward in such a calculation to ensure the cancellation of the electric current, except in the linear regime\cite{crepieux12} which is not the one we consider here.

Even if several theoretical works are devoted to the photo-assisted heat current in quantum dots\cite{rey07,chi12,arrachea07,caso10,caso11}, they are restricted to the calculation of its time average value which corresponds to the zero-order harmonic. A detailed study of higher order harmonics is still missing. The main objective of the present work is to fill this lack. This will allow us to know if the heat current through the junction will follow the time evolution of the modulated gate voltage, i.e., if it is sinusoidal, or if it has a more complicated profile. A careful study of the heat current and Peltier coefficient according to the amplitude and the frequency of the modulation, as done here, is thus needed.

The paper is organized as follows: in Sec.~II, we expose the model we have used to calculate the time-dependent heat current. In Secs.~III and IV, we discuss the photo-assisted heat current and the photo-assisted Peltier coefficient. Next, in Sec.~IV, we study in more details the harmonics of the ac heat current. We conclude briefly in Sec.~V.

\section{Model}

\begin{figure}[!ht]
\begin{center}
\includegraphics[width=7cm]{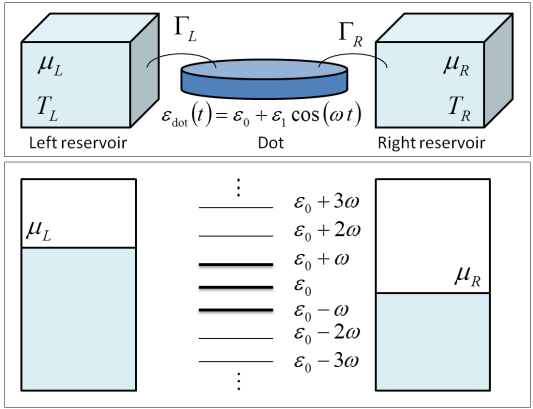}
\caption{(Upper panel) Schematic picture of the left (L) and right (R) reservoirs connected to a single level quantum dot with a modulated gate voltage. Notice that the definition we take for the currents means that their directions are oriented from each of the reservoirs to the dot. (Lower panel) Picture of the equivalent energy levels of the dot. The heat current is sinusoidal in the case where only the energy levels represented by a thick line contribute (i.e., the energy levels $\varepsilon_0$ and $\varepsilon_0\pm\omega$).
\label{figure1}}
\end{center} 
\end{figure}

We consider a single level non-interacting quantum dot connected to left (L) and right (R) reservoirs with chemical potentials $\mu_{L,R}$ at temperatures $T_{L,R}$ (see the upper panel of Fig.~\ref{figure1}). We define as well the source-drain voltage: $eV=\mu_L-\mu_R$, and the average temperature of the reservoirs: $T=(T_L+T_R)/2$. The energy level of the quantum dot can be modulated in time by tuning the gate voltage, we denote it $\varepsilon_{\mathrm{dot}}(t)$. We use the following Hamiltonian to describe this system:
\begin{eqnarray}
H=\sum_{k\in L,R}\varepsilon_k c_k^\dag c_k+\varepsilon_{\mathrm{dot}}(t)d^\dag d+\sum_{k\in L,R}V_kc_k^\dag d+h.c.~,\nonumber\\
\end{eqnarray}
where $\varepsilon_{k\in L,R}$ is the energy band of the reservoir L or R. The notations $c_k^\dag$ ($d^\dag$) and $c_k$ ($d$) refer to the creation and annihilation operators associated to the reservoirs (dot). $V_{k\in L,R}$ is the hopping amplitude between the reservoir L or R and the dot.

The time-dependent heat current through the left~(L) or right~(R) reservoirs, $I^h_{L,R}=-\dot Q_{L,R}$, is related to the energy current, $I^E_{L,R}=-\dot E_{L,R}$, and to the electric current, $I^e_{L,R}=-e\dot N_{L,R}$, by the identity:
\begin{eqnarray}\label{def_heat_current}
I^h_{L,R}(t)=I^E_{L,R}(t)-\frac{\mu_{L,R}}{e}I^e_{L,R}(t)~,
\end{eqnarray}
which is obtained from the thermodynamic relation: 
\begin{eqnarray}
dQ_{L,R}=dE_{L,R}-\mu_{L,R}dN_{L,R}~,
\end{eqnarray}
since the reservoirs are at equilibrium. The quantities $Q_{L,R}$, $E_{L,R}$ and $N_{L,R}$ correspond respectively to the heat, the energy and the number of particles in the left~(L) or right~(R) reservoirs. Thus, by calculating the electric and energy currents, one can deduce the heat current with the help of Eq.~(\ref{def_heat_current}). To determine these quantities, we use the fact that $i\hbar\dot N_{L,R}=[N_{L,R},H]$ and $i\hbar\dot E_{L,R}
=[E_{L,R},H]$ with:
\begin{eqnarray}
N_{L,R}&=&\sum_{k\in L,R}c_k^\dag c_k~,\\
E_{L,R}&=&\sum_{k\in L,R}\varepsilon_k c_k^\dag c_k~.
\end{eqnarray}

Notice that the definitions of the energies of the reservoirs, $E_{L,R}$, given above implicitly assume that we consider only the electronic contribution to the energy current (i.e., we neglect possible phononic contribution).

Within this model and with the help of Kelysh out-of-equilibrium Green's functions, the expression of the time-dependent heat current through the reservoir $p$ in the wide band limit reads as:\cite{crepieux11}
\begin{eqnarray}\label{time_heat_current}
&&I^h_p(t)=-\frac{\Gamma_{p}}{h}\bigg[2\int_{-\infty}^{\infty}(\varepsilon-\mu_p) f_{p}(\varepsilon)\mathrm{Im}\{A(\varepsilon,t)\} d\varepsilon\nonumber\\
&&+\sum_{p'=L,R}\Gamma_{p'}\int_{-\infty}^{\infty} (\varepsilon-\mu_{p})f_{p'}(\varepsilon)|A(\varepsilon,t)|^2 d\varepsilon
\bigg]~,
\end{eqnarray}
where $f_p$ is the Fermi-Dirac distribution function of the reservoir $p$. The enlargement of the dot energy level due to its coupling with the reservoir $p$ is defined as $\Gamma_{p}=2\pi\rho_p|V_p|^2$, where $\rho_p$ is the density of states of the reservoir $p$, and $V_p\equiv V_{k\in p}$. In the wide band limit, it is assumed to be energy independent. Notice that this is the energy current contribution, $I^E_p=-\dot E_p$, which is responsible of the presence of the $\varepsilon$ term in the heat current. 

The quantity $A(\varepsilon,t)$ in Eq.~(\ref{time_heat_current}) is the spectral function defined as:\cite{jauho94}
\begin{eqnarray}\label{spectral_function_definition}
A(\varepsilon,t)=\int_{-\infty}^{\infty}G^r(t,t_1)e^{i\varepsilon(t-t_1)/\hbar}dt_1~.
\end{eqnarray}

The retarded Green's function $G^r$ is the one of the dot connected to the reservoirs which is given here by:
\begin{eqnarray}\label{dot_green_function}
G^r(t,t_1)=g^r(t,t_1)e^{\Gamma(t_1-t)/2\hbar}~,
\end{eqnarray}
where $g^r(t,t_1)=-i\Theta(t-t_1)e^{-i\int_{t_1}^t dt_2\varepsilon_\mathrm{dot}(t_2)/\hbar}$ is the retarded Green's function of the isolated dot, $\Gamma=\Gamma_L+\Gamma_R$, and $\Theta$ is the Heaviside function.

In the present work, we consider the following time modulation for the dot energy level:
\begin{eqnarray}
\varepsilon_\mathrm{dot}(t)=\varepsilon_0+\varepsilon_1\cos(\omega t)~.
\end{eqnarray}
In that case, the spectral function of Eq.~(\ref{spectral_function_definition}) becomes:
\begin{eqnarray}\label{spectral_function_expression}
A(\varepsilon,t)&=&\sum_{n=-\infty}^{\infty}\sum_{m=-\infty}^{\infty}J_n\left(\frac{\varepsilon_1}{\hbar\omega}\right)J_m\left(\frac{\varepsilon_1}{\hbar\omega}\right)\frac{e^{i(n-m)\omega t}}{E_n(\varepsilon)+i\frac{\Gamma}{2}}~,\nonumber\\
\end{eqnarray}
where $E_n(\varepsilon)=\varepsilon-\varepsilon_0-n\omega$, and $J_n$ is the Bessel's function of order $n$ which appears in the expression of the spectral function through the identity $e^{ix\sin(\omega t)}=\sum_{n=-\infty}^{\infty}J_n(x)e^{in\omega t}$.

\section{Photo-assisted heat current}

In this section, we look at the time evolution of the heat current $I^h_p(t)$ which has been obtained by inserting Eq.~(\ref{spectral_function_expression}) in Eq.~(\ref{time_heat_current}) and integrating over energy numerically. We discuss its principal characteristics and compare them to the ones of the time-dependent electric current $I^e_p(t)$ that has a similar form than Eq.~(\ref{time_heat_current}) except the factors $(\varepsilon-\mu_p)$ which are not present.\cite{jauho94} 

\begin{figure}[!ht]
\begin{center}
\includegraphics[width=6cm]{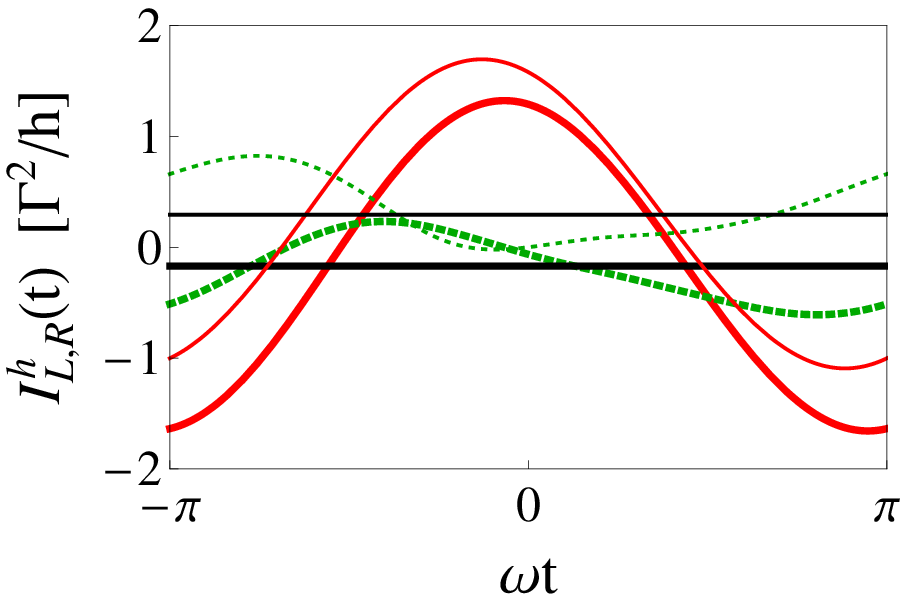}
\includegraphics[width=6cm]{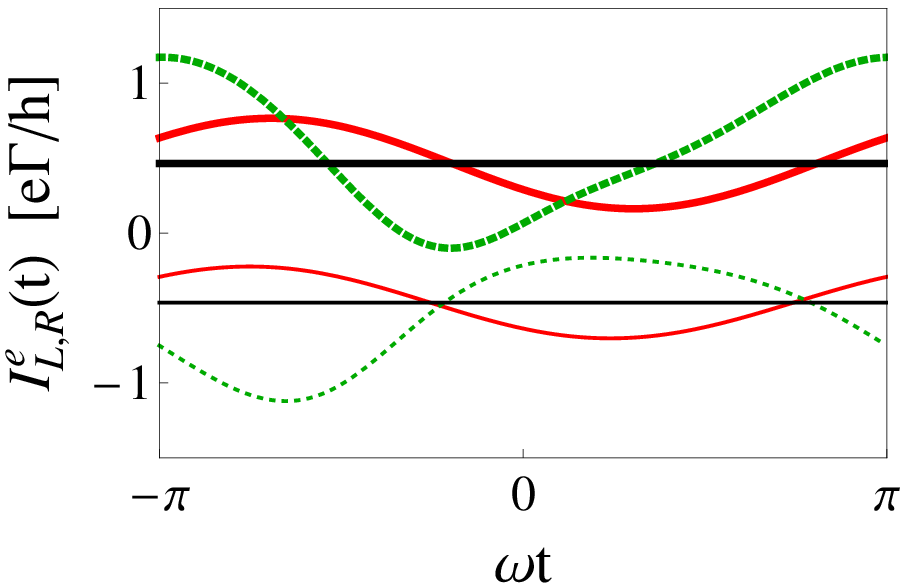}
\caption{(Upper panel) Heat current and (lower panel) electric current as a function of $\omega t$. The parameters are: $\varepsilon_0=1$, $\varepsilon_1=0.5$, $\Gamma_{L,R}=0.5$, $k_B T_{L,T}=0.01$ and $eV=1$. The modulation frequency is: $\hbar\omega=10$ (solid red lines) and $\hbar\omega=0.5$ (dotted green lines). The black straight lines correspond to the dc currents in the absence of modulation (i.e., $\varepsilon_1=0$). The thick (thin) lines correspond to the photo-assisted currents in the left $L$ (right $R$) reservoir. The unit for energies is $\Gamma$.
\label{figure2}}
\end{center} 
\end{figure}

\begin{figure}[!ht]
\begin{center}
\includegraphics[width=9cm]{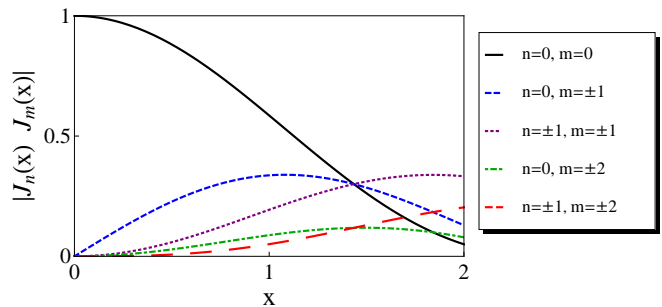}
\caption{Absolute value of the product $J_n(x)J_m(x)$ that appears in the spectral function of Eq.~(\ref{spectral_function_expression}) as a function of argument $x$ for different values of $n$ and $m$. When $x\lesssim 1$, this product decreases quickly when $n$ or $m$ increases. On the contrary, when $x>1$, there is no dominant terms in the sums over $n$ and $m$.
\label{figure3}}
\end{center} 
\end{figure}

It is important to notice that there are several characteristic energies in this system that are related to: the temperatures, $k_BT_{L,R}$, the source-drain voltage, $eV$, the modulation frequency, $\hbar\omega$, the amplitude of the modulation, $\varepsilon_1$, the dot energy level, $\varepsilon_0$, and the coupling strength between the dot and the reservoirs, $\Gamma$. The heat current depends on the relative values of all these energies. Here, we focus on the change in its behavior according to the value of the ratio $\varepsilon_1/\hbar\omega$ that appears in the argument of the Bessel's functions present in the expression of the heat current. Fig.~\ref{figure2} shows the time evolution of the heat and electric currents, at fixed $\varepsilon_1$ and temperatures $k_BT_{L,R}$ much smaller than all the other energies, for several values of the modulation frequency $\omega$ during a full period. We see that when $\hbar\omega\gg \varepsilon_1$, the left and right currents are both sinusoidal (see the solid red lines) and oscillate almost in phase around their corresponding dc values (see the black straight lines) whereas when $\hbar\omega\lesssim\varepsilon_1$, the signals are much more complex (see the dotted green lines). Indeed, when the ratio $\varepsilon_1/\hbar\omega$ is smaller than one, the dominant contributions in the heat current come from the terms associated with $n,m\in\{-1,0,1\}$ in the sums that appear in the spectral function of Eq.~(\ref{spectral_function_expression}) because the value of the product of Bessel's functions, $J_n(x)J_m(x)$, with argument $x$ smaller than one decreases quickly with increasing $n$ and $m$ (see Fig.~\ref{figure3}). As a consequence, we have:
\begin{eqnarray}\label{heat_current_appox}
I^{h}_p(t)\approx I^{h(0)}_p+2\mathrm{Re}\{I_p^{h(1)}e^{i\omega t}\}~,
\end{eqnarray}
where $I^{h (0)}_p$ is the zero-order harmonic (or average value) of the time-dependent heat current, and $I_p^{h (1)}$ its first order harmonic (see Sec.~V). This result explains the sinusoidal behavior we have obtained in the limit $\hbar\omega\gg\varepsilon_1$ (i.e., the red solid lines of Fig.~\ref{figure2}). On the contrary, when the ratio $\varepsilon_1/\hbar\omega$ is close or higher than one, a large number of terms in the sums over $n$ and $m$ in Eq.~(\ref{spectral_function_expression}) contribute and $I^{h}_p(t)$ is a superposition of several harmonics (see Sec.~V) which lead to a non sinusoidal signal (i.e., the green dotted lines of Fig.~\ref{figure2}).

\section{Photo-assisted Peltier coefficient}

We have all the ingredients to calculate the time evolution of the Peltier coefficient of the junction which is defined such as:\cite{crepieux11}
\begin{eqnarray}
\Pi(t)=\left.\frac{I_L^h(t)-I_R^h(t)}{I_L^e(t)-I_R^e(t)}\right|_{T_L=T_R}~.
\end{eqnarray}

By looking on Fig.~\ref{figure2}, we notice that whereas the left and right heat currents can be equal at some specific times (compare the thick and thin lines in the upper panel of Fig.~\ref{figure2}), this is not the case for the left and right electric currents which take distinct values at any time (compare the thick and thin lines in the lower panel of Fig.~\ref{figure2}). This property is general and not limited to the parameters we have chosen for plotting Fig.~\ref{figure2}. Thus, we can conclude that the Peltier coefficient of the junction will never be divergent. However, it vanish every times that the left and right heat currents are equal, i.e. $I_L^h(t)=I_R^h(t)$.

\begin{figure}[!ht]
\begin{center}
\includegraphics[width=6cm]{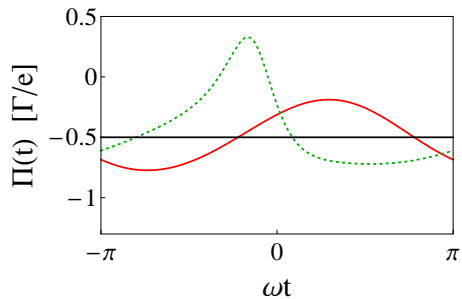}
\caption{Peltier coefficient as a function of $\omega t$. The parameters are the same to those of Fig.~\ref{figure2}: $\varepsilon_0=1$, $\varepsilon_1=0.5$, $\Gamma_{L,R}=0.5$, $k_B T_{L,T}=0.01$ and $eV=1$. The modulation frequency is: $\hbar\omega=10$ (solid red line), and $\hbar\omega=0.5$ (dotted green line). The black straight line corresponds to the Peltier coefficient in the absence of modulation (i.e., $\varepsilon_1=0$). The unit for energies is $\Gamma$.
\label{figure4}}
\end{center} 
\end{figure}

Figure \ref{figure4} shows the time evolution of the Peltier coefficient at low temperature (i.e., at $k_BT_{L,R}\ll \Gamma$). As it was the case for the heat and electric currents, the Peltier coefficient is sinusoidal in the limit $\hbar\omega\gg\varepsilon_1$ (see the solid red line): it follows the time evolution of the imposed gate voltage modulation. When the modulation frequency is reduced and becomes of the order of amplitude of the modulation $\hbar\omega\sim\varepsilon_1$, the Peltier coefficient is no longer sinusoidal (see the dotted green line).

Another quantity which characterizes the thermoelectricity is the Seebeck coefficient defined as the ratio between the voltage gradient and the temperature gradient in an open circuit: $S=\Delta V/\Delta T|_{I_{L,R}^e=0}$. However, this definition applies only in the linear response regime, as well as the direct relation between Peltier and Seebeck coefficients: $\Pi=ST$. Since here we are not necessarily in the linear response regime, the calculation of the photo-assisted Seebeck coefficient is a more difficult task which goes beyond the scope of the present study.

\section{Harmonics of the heat current}

We turn now our attention to the harmonics, $I_{L,R}^{h (N)}$, of the left and right heat currents which are defined through the relation:
\begin{eqnarray}\label{zero_fre_def}
I_{L,R}^h(t)=\sum_{N=-\infty}^{\infty}I_{L,R}^{h (N)}e^{iN\omega t}~.
\end{eqnarray}

The fact that $I_{L,R}^h(t)$ is a real quantity imposes that $I_{L,R}^{h (N)}$ and $I_{L,R}^{h (-N)}$ are complex conjugates.

Reporting Eq.~(\ref{spectral_function_expression}) in Eqs.~(\ref{time_heat_current}) and (\ref{zero_fre_def}), and performing the integration over time, we obtain:
\begin{eqnarray}\label{order_N}
&&I_{L,R}^{h (N)}=\frac{\Gamma_{L,R}}{2h}\sum_{\pm}\sum_{n=-\infty}^{\infty}J_n\left(\frac{\varepsilon_1}{\hbar\omega}\right)J_{n\pm N}\left(\frac{\varepsilon_1}{\hbar\omega}\right)\nonumber\\
&&\times\int_{-\infty}^{\infty}\Bigg\{\Big[\Gamma_{R,L}(f_{L,R}(\varepsilon)-f_{R,L}(\varepsilon))\mp 2if_{L,R}(\varepsilon)E_{n\pm N}(\varepsilon)\Big]\nonumber\\
&&\times\frac{\varepsilon-\mu_{L,R}}{(E_n(\varepsilon)\mp i\Gamma/2)(E_{n\pm N}(\varepsilon)\pm i\Gamma/2)}\Bigg\}d\varepsilon~,
\end{eqnarray}
where we have used the identity: $\sum_nJ_{n+m}(x)J_{n}(y)=J_{m}(x-y)$.

In the following, we study separately the zero-order harmonic of the heat current and the its harmonics of higher orders.

\subsection{Harmonic of order zero}

The zero-order ($N=0$) harmonic of the heat current corresponds to the time average heat current. From Eq.~(\ref{order_N}), we obtain:
\begin{eqnarray}\label{zero_freq_heat}
I_{L,R}^{h (0)}=\sum_{n=-\infty}^{\infty}J_n^2\left(\frac{\varepsilon_1}{\hbar\omega}\right)I_{L,R}^{h (dc)}(\varepsilon_0+n\omega)~,
\end{eqnarray}
where $I_{L,R}^{h (dc)}$ is the dc heat current defined as:
\begin{eqnarray}\label{definition_dc_current}
I_{L,R}^{h (dc)}(z)=\frac{\Gamma_L\Gamma_R}{h}\int_{-\infty}^{\infty}\frac{(\varepsilon-\mu_{L,R})(f_{L,R}(\varepsilon)-f_{R,L}(\varepsilon))}{(\varepsilon-z)^2+\Gamma^2/4}d\varepsilon~.\nonumber\\
\end{eqnarray} 

The description of Eq.~(\ref{zero_freq_heat}) is the following: the zero-order harmonic of the ac heat current is equal to an infinite sum over $n$ of dc heat currents associated to shifted dot energies: $\varepsilon_0+n\omega$, times the square of the Bessel's function of order $n$ (see the lower panel of Fig.~\ref{figure1} for a schematic representation of those shifted energy levels). A similar kind of relation linking the zero-order harmonic of the electric current, $I_p^{e (0)}$, and dc electric currents, $I_{p}^{e (dc)}(\varepsilon_0+n\omega)$, holds\cite{tucker85}:
\begin{eqnarray}
I_{L,R}^{e (0)}=\sum_{n=-\infty}^{\infty}J_n^2\left(\frac{\varepsilon_1}{\hbar\omega}\right)I_{L,R}^{e (dc)}(\varepsilon_0+n\omega)~,
\end{eqnarray}
where:
\begin{eqnarray}\label{definition_dc_current_el}
I_{L,R}^{e (dc)}(z)=\frac{\Gamma_L\Gamma_R}{h}\int_{-\infty}^{\infty}\frac{f_{L,R}(\varepsilon)-f_{R,L}(\varepsilon)}{(\varepsilon-z)^2+\Gamma^2/4}d\varepsilon~.
\end{eqnarray}

Even if the expressions of the zero-order harmonics of the heat and electric currents look similar, i.e. they contain an infinite sum of dc contributions, they differ on an essential point: whereas $I^{e (0)}_L=-I^{e (0)}_R$, we have $I^{h (0)}_L\ne-I^{h (0)}_R$ because of the $(\varepsilon-\mu_{L,R})$ factor which is present in Eq.~(\ref{definition_dc_current}) but not in Eq.~(\ref{definition_dc_current_el}). This leads to rather different profile for the zero-order harmonics of the heat and electric currents as detailed below.

\begin{figure}[!ht]
\begin{center}
\includegraphics[width=6cm]{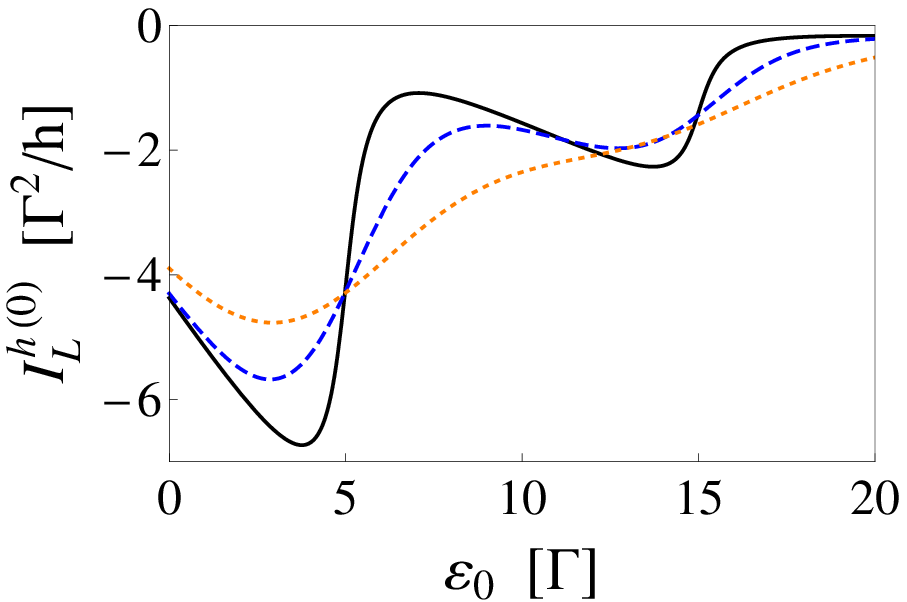}
\includegraphics[width=6cm]{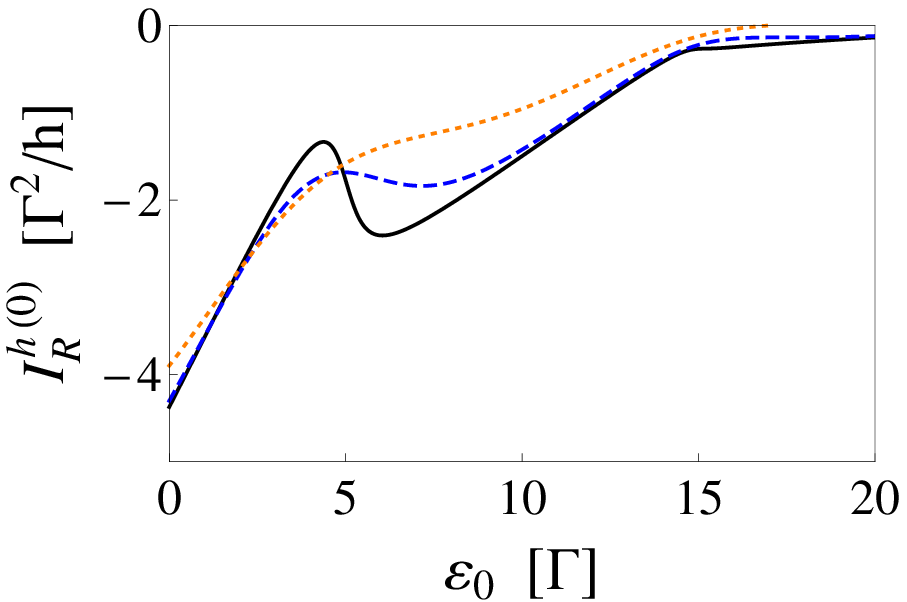}
\caption{Zero-order harmonic of the left heat current $I^{h (0)}_L$ (upper panel) and right heat current $I^{h (0)}_R$ (lower panel) as a function of $\varepsilon_0$. The temperatures are $k_BT_{L,T}=0.01$ (solid black lines), $k_BT_{L,T}=1$ (dashed blue lines) and $k_BT_{L,T}=2$ (dotted orange lines). The other parameters are $\varepsilon_1=10$, $eV=-10$, and $\hbar\omega=10$. The unit for energies is $\Gamma$.
\label{figure5}}
\end{center} 
\end{figure}

\begin{figure}[!ht]
\begin{center}
\includegraphics[width=6cm]{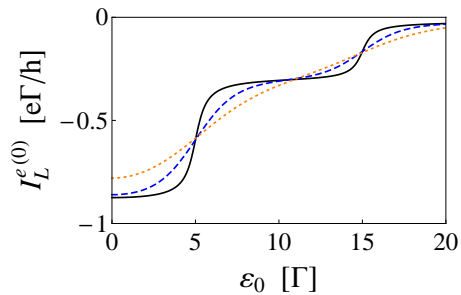}
\caption{Zero-order harmonic of the electric current $I^{e (0)}_L=-I^{e (0)}_R$ as a function of $\varepsilon_0$. The parameters and the legends are the same to those of Fig.~\ref{figure5}. The unit for energies is $\Gamma$.
\label{figure6}}
\end{center} 
\end{figure}

Figure \ref{figure5} shows the zero-order harmonic of the left and right heat currents at fixed frequency modulation, $\hbar\omega=10\Gamma$ and voltage, $eV=-10\Gamma$, as a function of $\varepsilon_0$. We observe a jump each time that the energy $(\varepsilon_0\pm V/2)$ is an integer multiple of $\hbar\omega$, here: $\varepsilon_0=5\Gamma$, $\varepsilon_0=15\Gamma$, etc... Between these points, the left (right) heat current slowly decreases (increases) contrary to what is observed for electric current which takes a constant value between the jumps (see Fig.~\ref{figure6}). The staircase behavior of the zero-order harmonic of the electric current is a standard result\cite{lesovik94} in the regime corresponding to the one of Figs.~\ref{figure5} and \ref{figure6} (i.e., $\Gamma\ll\{\hbar\omega,\varepsilon_1,eV\}$). It comes from the fact that the dot occupation is a conserved quantity in the stationary regime, contrary to the dot heat which is not conserved, even in the stationary regime, because of the energy dissipation within the tunnel barriers\cite{crepieux11,crepieux12}.
 
The width of the jumps depends both on the temperatures and on the coupling strength between the quantum dot and the reservoirs (i.e., on the value of $\Gamma$). The higher the temperatures, the larger the jump width (compare the solid black, dashed blue and dotted orange lines in Figs.~\ref{figure5} and \ref{figure6}). The increasing of the coupling $\Gamma$ enhances also the larger of the jump width and when $\Gamma\sim\{\hbar\omega,\varepsilon_1,eV\}$, the staircase structure is not more visible (not shown).

\subsection{Harmonic of order $N$}

\begin{figure}[!ht]
\begin{center}
\includegraphics[width=6cm]{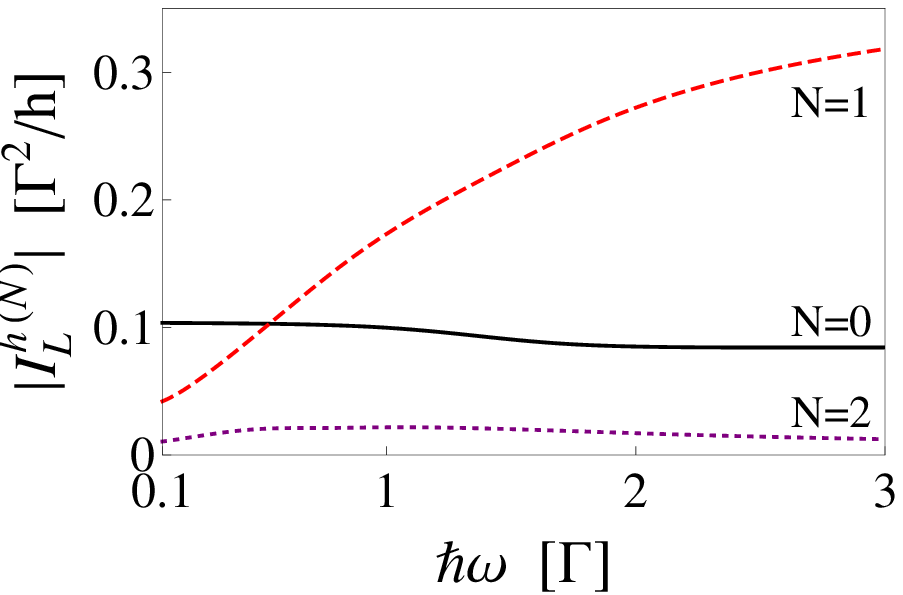}
\includegraphics[width=6cm]{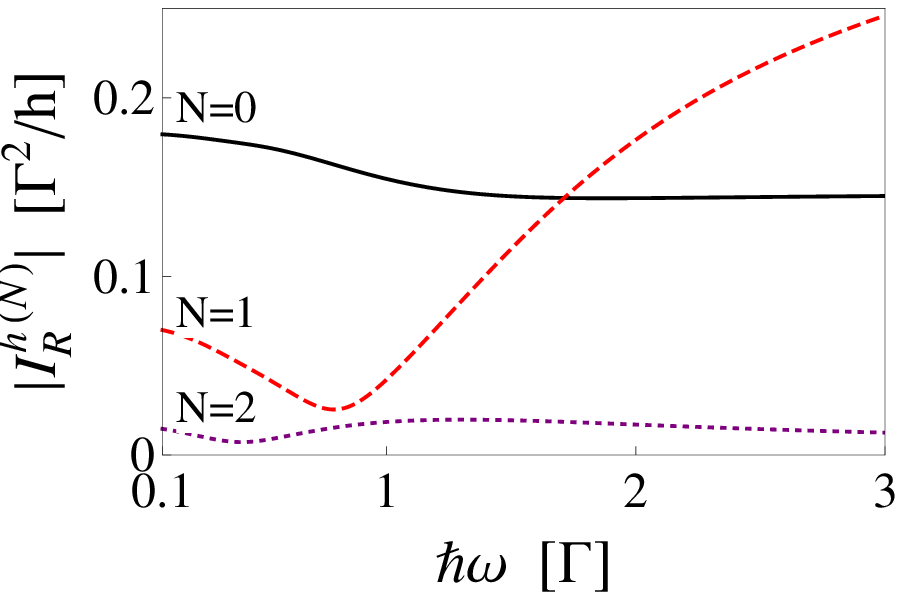}
\caption{Amplitude of the zero-order (solid black lines), first order (dashed red lines) and second order (dotted purple lines) harmonics of the left heat current (upper panel) and right heat current (lower panel). The parameters are $k_BT_{L,R}=0.01$, $\varepsilon_0=1$, $eV=1$, and $\varepsilon_1=0.5$. The unit for energies is $\Gamma$.
\label{figure7}}
\end{center} 
\end{figure}

With the help of Eq.~(\ref{order_N}), we have calculated numerically the harmonics of the heat currents. In Fig.~\ref{figure7} is plotted the module of the harmonics of order $0$, $1$ and $2$ as a function of the modulation frequency. The module of the higher order harmonics of the heat currents, i.e. $|I_{L,R}^{h(N>2)}|$, are not shown on the graphs because they are much smaller in amplitude in comparison to $|I_{L,R}^{h(2)}|$.

A crucial information that we extract from Fig.~\ref{figure7} is the fact that when $\hbar\omega\lesssim\varepsilon_1$, the modules of first and second harmonics are of the same order in magnitude: $|I_{L,R}^{h(1)}|\sim|I_{L,R}^{h(2)}|$, whereas when $\hbar\omega\gg\varepsilon_1$, the first harmonic dominates over the second harmonic (compare the dashed red lines to the dotted purple lines). These last result justifies the approximation that we have made in order to write Eq.~(\ref{heat_current_appox}) and explain the sinusoidal behavior of the time-dependent heat current obtained in that regime.

\section{Conclusion}

We have highlighted several interesting features in the photo-assisted heat current through a metal/dot/metal junction: (i) its zero-order harmonic exhibits a devil staircase with non horizontal steps which result from the fact that the dot heat in not a conserved quantity even in the stationary regime because of energy dissipation within tunnel barriers; (ii) the time evolution of the heat current is non sinusoidal when the modulation frequency of the gate voltage is of the order of the amplitude of the modulation; and (iii) the time evolution of the associated Peltier coefficient is strongly dependent of the energy scales which characterize the junction.

%%%%%%%%%%%%%%%%%%%%%%%%%%%%%%%%%%%%%%%%%%%%%%%%%%%%%%%%%%%%%%

\end{document}